\documentclass[twocolumn,english,pra,aps,eqsecnum,longbibliography]{revtex4-2}
\usepackage[LGR,T1]{fontenc}
\usepackage[latin9]{inputenc}
\usepackage{color}
\usepackage{bm}
\usepackage{amsmath}
\usepackage{amssymb}

\makeatletter

\providecommand{\tabularnewline}{\\}

\makeatother

\usepackage{babel}
\begin{document}
\title{Interacting fermion dynamics in Majorana phase-space}
\author{Ria~Rushin~Joseph }
\affiliation{Centre for Quantum Technology Theory, Swinburne University of Technology,
Melbourne 3122, Australia}
\author{Laura~E~C~Rosales-Zárate}
\affiliation{Centro de Investigaciones en Óptica A.C., León, Guanajuato 37150,
México}
\author{Peter~D~Drummond }
\affiliation{Centre for Quantum Technology Theory, Swinburne University of Technology,
Melbourne 3122, Australia}
\begin{abstract}
The problem of fermion dynamics is studied using the Q-function for
fermions. This is a probabilistic phase-space representation, which
we express using Majorana operators, so that the phase-space variable
is a real antisymmetric matrix. We consider a general interaction
Hamiltonian with four Majorana operators and arbitrary properties.
Our model includes the Majorana Hubbard and Fermi Hubbard Hamiltonians,
as well as general quantum field theories of interacting fermions.
Using the Majorana Q-function we derive a generalized Fokker-Planck
equation, with results for the drift and diffusion terms. The diffusion
term is proved to be traceless, which gives a dynamical interpretation
as a forwards-backwards stochastic process. This approach leads to
a model of quantum measurement in terms of an ontology with real vacuum
fluctuations.
\end{abstract}
\maketitle

\section{\label{sec:level1}Introduction}

Fermionic physics is universal: all stable massive elementary particles
are fermions. Here we analyze the nonlinear dynamics of a probabilistic
phase-space of fermions \citep{Husimi1940,Ria:2018_Majorana}. Recent
developments in quantum measurement theory \citep{PM2020_RetrocausalQField}
have led to a theory of bosonic quantum dynamics as stochastic processes
in phase-space, propagating in both time directions. We obtain the
time-evolution equation for interacting fermions in Majorana phase-space,
and show that it is also generalized Fokker-Planck equation with a
traceless diffusion matrix \citep{Altland_PRL2012_Qchaos}. This demonstrates
that retrocausal physics \citep{dirac1938classical,wharton2020colloquium}
of fermions is equivalent to quantum mechanics.

Such phase-space methods are potentially relevant to other developments
in quantum computing \citep{nayak2008non}, and to the control acquired
in ultra-cold Fermi systems, which allows studies of strongly interacting
fermions~\citep{Bloch:2012,Bloch_RMP_2008_ManyBodyPhys_UG}. Experiments
in this area include superfluidity~\citep{Greiner:2003,Zwierlein:2005},
the crossover from Cooper pairs to the Bose-Einstein condensate (BEC)
region~\citep{Greiner:2003,Jochim:2003,Regal:2004,Zwierlein:2005},
and the Hubbard model~\citep{Hubbard_1963,Bloch_2005,Jaksch:2005,Esslinger:2010,Hofstetter:2018},
which has been realized in optical lattices~\citep{Bloch_2005,Jaksch:2005,Lewenstein:2007}.
This exhibits phase-transitions~\citep{Kohl:2005,Greiner:2002PT,hofrichter2016direct,Greiner_2016},
transport properties~\citep{Strohmaier:2007,Schneider:2012}, and
anti-ferromagnetism~\citep{hart2015observation,hilker2017revealing,boll2016spin,fratino2017signatures}.
It can be used as a quantum simulator~\citep{Hofstetter:2018,Tarruel:2018}.

Since interactions play an important role in Fermi systems, it is
important to develop first-principles theoretical methods to investigate
the corresponding dynamics, without mean-field approximations. Some
of the different theoretical methods that have been used to study
strongly interacting fermions~\citep{Giorgini_RMP2008_Theory_UltracoldFG,Liu:2013,VanHoucke:2012}
include Monte-Carlo methods, which have a sign problem~\citep{Troyer:2005},
and Grassmann phase-space approaches \citep{Cahill_Glauber_1969_Density_operators,Gardiner_Book_QNoise,dalton2016grassmann},
although these can become exponentially complex. 

An alternative approach is via Gaussian phase-space representations~\citep{Corney_PD_JPA_2006_GR_fermions,Corney_PD_PRB_2006_GPSR_fermions}
and a fermionic P-function~\citep{Corney_PD_PRL2004_GQMC_ferm_bos,Corney_PD_PRB_2006_GPSR_fermions,Corboz_Chapter_PhaseSpaceMethodsFermions}.
Such methods have been used to investigate the Fermi-Hubbard model~\citep{Corney_PD_PRL2004_GQMC_ferm_bos,Corney_PD_PRB_2006_GPSR_fermions,Imada_2007_GBMC,Imada:2007_2DHM_Superconductivity,Corboz_Chapter_PhaseSpaceMethodsFermions},
but can lead to sampling errors \citep{Assaad:2005_ProjectedGMCM,Corboz_2008_Projection_GR}.
The technique used here is the generalized Q-function~\citep{FermiQ}
defined in terms of Gaussian operators. Rather than using Fermi ladder
operators, we use Majorana operators. This corresponds to the Majorana
phase-space~\citep{Ria:2018_Majorana}, which has been used to study
the dynamics of shock waves~\citep{Riashock2018} and information-related
quantities like the Renyi entropy, purity and fidelity~\citep{riaentropy}.
Here we extend this to fermion interactions. Interactions with bosonic
fields have been treated elsewhere \citep{drummond2021time}.

Majorana fermions and their related group structures have been heavily
investigated in their own right in recent years~\citep{Rahmani:2019MajRev,chiu2015strongly}.
 One of their features is that they are their own antiparticle~\citep{Majorana:1937}
and are found in topological superconductors~\citep{Beenakker:2015,Sato:2016}.
They have a possible role in quantum computation~\citep{nayak2008non,akhmerov2010topological,alicea2011non,DasSarma:2015,vijay2015majorana,OBrien:2018}
as well as more generally in condensed matter physics~\citep{Mourik:2012,benedek2020majorana,motome2020hunting,nayak2008non,Elliott:2015,Beenaker:2013},
due to their relationship with the general $DIII$ symmetry class
\citep{altland1997nonstandard}. One major topic of study is to incorporate
interactions ~\citep{Rahmani:2019MajRev}. There are several Majorana
models that include interactions, including the Majorana Hubbard model
on square lattices~\citep{Affleck_Rahmani_Pikulin_2017,Rahmani:2019MajRev,chiu2015strongly},
honeycomb lattices~\citep{dutreix2014majorana,lahtinen2011interacting,li2018majorana}
triangular lattices~\citep{kraus2011majorana} and vortex lattices~\citep{biswas2013majorana}. 

The study of interactions for Majorana fermions in condensed matter
physics~\citep{Alicea:2012}, leads to a new classification of topological
phases in one dimension~\citep{Fidkowski:2011}. We treat the most
general interactions of four operators, using a Majorana Q-function~\citep{Ria:2018_Majorana}.
This includes the Fermi Hubbard and other lattice models of quartic
fermion interactions ~\citep{Affleck_Rahmani_Pikulin_2017,Rahmani:2019MajRev}.
In this paper, we obtain the generalized Fokker-Planck equation that
describes Q-function dynamics. Our approach uses a phase-space of
real antisymmetric matrices, which is a fundamental concept in group
theory \citep{Cartan:1935}. This is related to variational theories
of Gaussian states \citep{hackl2020geometry}, except that Q-functions
do not require a variational approximation. 

This paper is organized as follows: Section \ref{sec:Gaussian-Majorana-Q-function}
gives a summary of Majorana Q-functions and notation. In Section \ref{sec:Hamiltonian-of-the}
we discuss the Hamiltonian and dynamical evolution. Sec. \ref{sec:The-Fokker-Planck}
gives a derivation of the Fokker Planck equation, and properties of
the diffusion term. In Sec. \ref{sec:DriftTermPSD} we discuss the
relation of the drift to the phase space of pure states. A summary
is given in Sec. \ref{sec:Summary}. 

\section{Gaussian Majorana Q-function\label{sec:Gaussian-Majorana-Q-function}}

Phase-space methods have been used to study bosonic fields \citep{Carter:1987,Drummond_EPL_1993,Corney_2008,Blakie:2008,Polkovnikov_2010_Phase_Rep_QDyn,He2012_QDynamics_FrontiersPhysics}
with considerable success in comparisons to experiment \citep{Drummond1993_Nature,Corney:2006_ManyBodyQD}.
One can analogously define fermionic phase-space representations using
Grassmann variables~\citep{Cahill_Glauber_fermions_1999,dalton2016grassmann}.
However these non-commuting variables have an exponential complexity. 

Fermionic phase-space representations have also been introduced over
complex phase-spaces, including the P-function~\citep{Corney_PD_PRL2004_GQMC_ferm_bos,Corney_PD_JPA_2006_GR_fermions,Corney_PD_PRB_2006_GPSR_fermions}
and the Q-function~\citep{FermiQ}. These use ordered Gaussian fermionic
operators as a basis, together with a complex phase-space. 

Here we treat a third approach, a representation which uses as a phase-space
variable a real anti-symmetric matrix~\citep{Ria:2018_Majorana}.
The advantage of this approach is that it corresponds mathematically
to a well-defined compact homogeneous space. This representation is
the Majorana Q-function. We start by giving a brief summary of its
properties. 

\subsection{Gaussian operator definition}

We consider a general $M$-mode lattice of quantum fermionic modes
described by $M$ fermionic annihilation and creation operators $\hat{\bm{a}}$,
$\hat{\bm{a}}^{\dagger}$. We denote $M$-dimensional vectors and
matrices with a bold notation, and $2M$ dimensional vectors and matrices
with an underline, so that an extended operator $\hat{\underline{a}}$
is defined as
\begin{equation}
\hat{\underline{a}}=\left(\hat{a}_{1}\ldots\hat{a}_{M},\hat{a}_{1}^{\dagger}\ldots\hat{a}_{M}^{\dagger}\right)^{T}.
\end{equation}
One can obtain $2M$-dimensional Majorana operators from this extended
vector of fermionic creation and annihilation operators~\citep{Corney_PD_PRB_2006_GPSR_fermions}
by the action of a matrix~\citep{Balian_Brezin_Transformations},
$\underline{\underline{U_{0}}}=\left[\begin{array}{cc}
\mathbf{I} & \mathbf{I}\\
-i\mathbf{I} & i\mathbf{I}
\end{array}\right]$, so that:
\begin{equation}
\hat{\underline{\gamma}}=\underline{\underline{U_{0}}}\hat{\underline{a}}.
\end{equation}

With this relation, $\hat{\gamma}_{1}=\hat{a}_{1}+\hat{a}_{1}^{\dagger}$,
and $\hat{\gamma}_{M+1}=i\left(\hat{a}_{1}^{\dagger}-\hat{a}_{1}\right)$.
The resulting real Majorana operator $\hat{\underline{\gamma}}$ is
a $2M-$dimensional Hermitian Fermi operator which obeys the following
anti-commutation relation, for $i,j=1,\ldots2M$:
\begin{equation}
\left\{ \hat{\gamma_{i}},\hat{\gamma_{j}}\right\} =2\delta_{ij}.
\end{equation}

The Gaussian Majorana operator can be defined in several ways, either
ordered or unordered. To obtain operator differential identities,
we choose a normal ordering approach. The phase-space variable is
defined for general Gaussian operators using an antisymmetric complex
matrix, $\underline{\underline{x}}$, in one of the irreducible bounded
symmetric domains of group theory, defined \citep{Cartan:1926,Cartan:1927,Cartan:1935,Hua_Book_harmonic_analysis}
so that
\begin{equation}
\underline{\underline{x}}\underline{\underline{x}}^{\dagger}\le\underline{\underline{I}}.
\end{equation}
Our definition of a Gaussian basis gives an exponential of a quadratic
in the Majorana operators as:
\begin{equation}
\hat{\Lambda}\left(\underline{\underline{x}}\right)=N\left(\underline{\underline{x}}\right):\exp\Biggl[-i\hat{\underline{\gamma}}^{T}\left[\underline{\underline{i}}+\left(\underline{\underline{i}}+\underline{\underline{i}}\underline{\underline{x}}\underline{\underline{i}}\right)^{-1}\right]\hat{\underline{\gamma}}/2\Biggr]:.\label{eq:MajOpX}
\end{equation}
Here $N\left(\underline{\underline{x}}\right)$ ensures that the Gaussian
operator is normalized so that $\mathrm{Tr}\left[\hat{\Lambda}\left(\underline{\underline{x}}\right)\right]=1$,
and we define $\underline{\underline{i}}=\left[\begin{array}{cc}
\mathbf{0} & \mathbf{I}\\
-\mathbf{I} & \mathbf{0}
\end{array}\right]$, which is a matrix square root of $-\underline{\underline{I}}$ .

From now on, we treat the case where the Gaussian operator is Hermitian
and positive definite, so that $\underline{\underline{x}}$ is a real
anti-symmetric matrix, with fermionic pure states occurring at the
boundary where $\underline{\underline{x}}\underline{\underline{x}}^{T}=\underline{\underline{I}}$.
Gaussian states on the boundary have also been classified as a fundamental
symmetry class in the physics literature corresponding to certain
condensed matter devices \citep{Altland_Zirnbauer:1997}. 

\subsection{Q-function definition}

Q-function phase-space representations are positive probability distributions
that can provide powerful simulation methods. These include, for example,
a recent 60 qubit simulation of mesoscopic multipartite Bell violations
\citep{ReidqubitPhysRevA.90.012111}, using GHZ states in ion traps.
A Majorana Q-function can be defined for any fermionic quantum density
matrix, $\hat{\rho}$. This is defined as a distribution over the
antisymmetric matrices:
\begin{equation}
Q\left(\underline{\underline{x}}\right)=\mathrm{Tr}\left[\hat{\rho}\hat{\Lambda}^{N}\left(\underline{\underline{x}}\right)\right],\label{eq:MajQf}
\end{equation}
where we introduce $\hat{\Lambda}^{N}\left(\underline{\underline{x}}\right)$
as a rescaling of the unit trace Gaussian $\hat{\Lambda}\left(\underline{\underline{x}}\right)$,
such that:
\begin{equation}
\hat{\Lambda}^{N}\left(\underline{\underline{x}}\right)=\frac{1}{\mathcal{N}}\hat{\Lambda}\left(\underline{\underline{x}}\right)S\left(\underline{\underline{x}}^{2}\right).
\end{equation}

The function $S\left(\underline{\underline{x}}^{2}\right)$ is an
arbitrary even function of $\underline{\underline{x}}$, and the normalization
$\mathcal{N}$ is defined so that the following resolution of identity
holds, 
\begin{equation}
\hat{1}=\int d\underline{\underline{x}}\hat{\Lambda}^{N}\left(\underline{\underline{x}}\right),
\end{equation}
where the antisymmetric real matrix integration measure \citep{Hua_Book_harmonic_analysis}
is given by 
\begin{equation}
d\underline{\underline{x}}=\prod_{1\le j<k\le2M}dx_{ij}.
\end{equation}
As a result, since $\mathrm{Tr}\left[\hat{\rho}\right]=1$, the probability
distribution is normalized to unity
\begin{equation}
\int d\underline{\underline{x}}Q\left(\underline{\underline{x}}\right)=1.
\end{equation}

Any fermionic observable can be calculated using the fermionic $Q$-function,
together with the appropriate identities. To give an example, the
expectation value of the Majorana two-fermion correlation function,
\begin{equation}
\hat{X}_{\mu\nu}\equiv\frac{i}{2}\left[\gamma_{\mu},\gamma_{\nu}\right]\label{eq.Xcape}
\end{equation}
 is given by~\citep{Ria:2018_Majorana}:
\begin{equation}
\Bigl\langle\underline{\underline{\hat{X}}}\Bigr\rangle=\left(4M-1\right)\int\underline{\underline{x}}Q\left(\underline{\underline{x}}\right)d\underline{\underline{x}}.
\end{equation}

Explicit results obtained here will use the limit of $S\left(\underline{\underline{x}}^{2}\right)=1$
for simplicity. Other choices are also possible, including the pure
states with $\underline{\underline{x}}^{2}=-\underline{\underline{I}}$,
which are divided into two parity classes \citep{riaentropy}, each
belonging to the DIII symmetric space of Cartan \citep{altland1997nonstandard}.

\subsection{Notation and derivatives\label{subsec:usefulderivatives}}

In this section we give a summary of the derivatives that will be
used, as well as the compact notation that is introduced. The phase-space
variable of the Majorana Q-function is a real antisymmetric matrix.
Derivatives with respect to $x$ are defined to take account of this
constraint, i.e. so that $x_{ab}\equiv-x_{ba}$ \citep{Ria:2018_Majorana},
so that:
\begin{equation}
\frac{\partial x_{ab}}{\partial x_{cd}}\equiv\delta_{ac}\delta_{bd}-\delta_{ad}\delta_{bc}.\label{eq:deranti}
\end{equation}
We note that antisymmetric derivatives $d/d\underline{\underline{x}}$
are defined here so that $\left(\partial/\partial\underline{\underline{x}}\right)_{ij}\equiv\partial_{ji}=-\partial_{ij}$.
The differential identities given below are given in terms of the
matrices $\underline{\underline{x}}^{\pm}\equiv\underline{\underline{x}}\pm i\underline{\underline{I}}.$
As a result,
\begin{equation}
\frac{\partial x_{ab}^{+}}{\partial x_{cd}}=\frac{\partial x_{ab}}{\partial x_{cd}}=\frac{\partial x_{ab}^{-}}{\partial x_{cd}},
\end{equation}
since $\delta_{ab}$ is a constant. We also define:
\begin{equation}
\partial_{\alpha\beta}\equiv\frac{\partial}{\partial x_{\alpha\beta}}.
\end{equation}
Where appropriate, we will use a shorthand form with a restricted
index range that only includes the independent parameters, where $\bm{\alpha}\equiv\left(\alpha,\beta\right)$
with $1\le\alpha<\beta\le2M$, similarly $\bm{\mu}=\left(\mu,\nu\right)$,
and hence, for $\alpha<\beta$
\begin{equation}
\partial_{\bm{\alpha}}\equiv\partial_{\alpha\beta}\equiv\frac{\partial}{\partial x_{\alpha\beta}}.
\end{equation}

Derivatives of products of anti-symmetric matrices can be obtained
by using the product rule, for example:
\begin{align}
\partial_{cd}\left(x_{ab}x_{ef}\right) & =x_{ab}\left(\delta_{ce}\delta_{df}-\delta_{cf}\delta_{de}\right)\nonumber \\
 & +x_{ef}\left(\delta_{ac}\delta_{bd}-\delta_{ad}\delta_{bc}\right).\label{eq:productrule}
\end{align}

The indices of $\underline{\underline{x}}^{+}$ and $\underline{\underline{x}}^{-}$
are related through the following expression:
\begin{equation}
x_{ab}^{+}=-x_{ba}^{-}.\label{eq:inver}
\end{equation}
Throughout the paper we use a four-index notation for the products
of the form $\underline{\underline{x}}^{\pm}\underline{\underline{x}}^{\mp}$,
which is:
\begin{align}
X_{ij}^{\alpha\beta} & \equiv x_{i\alpha}^{+}x_{\beta j}^{-},\nonumber \\
X_{ij}^{\alpha\beta*} & \equiv x_{i\alpha}^{-}x_{\beta j}^{+}.\label{eq:x-x+}
\end{align}

From Eq. (\ref{eq:inver}), one obtains:
\begin{align}
X_{ji}^{\beta\alpha} & =X_{ij}^{\alpha\beta}.\label{eq:inver2}
\end{align}
 Real and imaginary parts are given by
\begin{eqnarray}
\Im X_{ij}^{\alpha\beta} & = & -i\left(X_{ij}^{\alpha\beta}-X_{ij}^{\alpha\beta*}\right)/2\nonumber \\
 & = & -\left(x_{i\alpha}\delta_{\beta j}+\delta_{i\alpha}x_{j\beta}\right),\nonumber \\
\Re X_{ij}^{\alpha\beta} & = & \left(X_{ij}^{\alpha\beta}+X_{ij}^{\alpha\beta*}\right)/2\nonumber \\
 & = & -\left(x_{i\alpha}x_{j\beta}-\delta_{i\alpha}\delta_{\beta j}\right).\label{eq:ReXij+}
\end{eqnarray}
Derivatives of the variable $X_{ij}^{\alpha\beta}$ are calculated
using the product rules and the definition of the derivative given
in Eq. (\ref{eq:deranti}), for example:
\begin{align}
\partial_{\mu\nu}X_{ij}^{\alpha\beta} & =\partial_{\boldsymbol{\mu}}X_{ij}^{\bm{\alpha}}\label{eq:Xabij}\\
 & =x_{j\beta}^{+}\left(\delta_{\alpha\mu}\delta_{i\nu}-\delta_{\alpha\nu}\delta_{i\mu}\right)+x_{i\alpha}^{+}\left(\delta_{\beta\mu}\delta_{j\nu}-\delta_{\beta\nu}\delta_{j\mu}\right).\nonumber 
\end{align}

Some useful derivatives used in the calculations are:
\begin{eqnarray}
\left(\partial_{\bm{\alpha}}x_{kl}^{+}\right)X_{kl}^{\bm{\alpha}} & = & X_{ij}^{kl}-X_{ij}^{lk}.\nonumber \\
\left(\partial_{\bm{\alpha}}x_{ij}^{+}\right)X_{kl}^{\bm{\alpha}*} & = & X_{kl}^{ij*}-X_{kl}^{ji*},\nonumber \\
\partial_{\bm{\alpha}}\Im X_{ij}^{\bm{\alpha}} & = & 0,\nonumber \\
\partial_{\bm{\alpha}}\Re X_{ij}^{\bm{\alpha}} & = & -2x_{ij}\left(2M-1\right).\label{eq:Der}
\end{eqnarray}

\subsection{Majorana differential identities\label{subsec:Majorana-differential-identities}}

The utility of the normally ordered approach is that straightforward
differential identities exist for all fermionic observables. Observables
are even polynomials in Majorana operators, and their identities can
be obtained from the quadratic results given here. These are essential
in order to obtain dynamical equations of motion and observables using
phase-space representations. 

For the Majorana Gaussian operator, quadratic differential identities
were derived in~\citep{Ria:2018_Majorana}, and are given below.\\

\begin{itemize}
\item Left product:
\begin{equation}
\hat{\underline{\gamma}}\hat{\underline{\gamma}}^{T}\hat{\Lambda}=i\left[\underline{\underline{x}}^{-}\frac{d\hat{\Lambda}}{d\underline{\underline{x}}}\underline{\underline{x}}^{+}-\hat{\Lambda}\underline{\underline{x}}^{+}\right].\label{eq:UnorderedDifId}
\end{equation}
\item Right product:
\begin{equation}
\hat{\Lambda}\hat{\underline{\gamma}}\hat{\underline{\gamma}}^{T}=i\left[\underline{\underline{x}}^{+}\frac{d\hat{\Lambda}}{d\underline{\underline{x}}}\underline{\underline{x}}^{-}-\hat{\Lambda}\underline{\underline{x}}^{+}\right].\label{eq:UnorderedDifId-1}
\end{equation}
\item Mixed product:
\begin{equation}
\hat{\underline{\gamma}}\hat{\Lambda}\hat{\underline{\gamma}}^{T}=i\left[-\underline{\underline{x}}^{-}\frac{d\hat{\Lambda}}{d\underline{\underline{x}}}\underline{\underline{x}}^{-}+\hat{\Lambda}\underline{\underline{x}}^{-}\right].\label{eq:unmixed}
\end{equation}
\item Commutator product:
\begin{equation}
\left[\gamma_{i}\gamma_{j}-\gamma_{j}\gamma_{i},\widehat{\Lambda}\right]=4\left[x_{\kappa j}\frac{d\hat{\Lambda}}{dx_{\kappa i}}-x_{i\kappa}\frac{d\hat{\Lambda}}{dx_{j\kappa}}\right].\label{eq:COMMU}
\end{equation}
\end{itemize}
Here $\underline{\underline{x}}^{\pm}\equiv\underline{\underline{x}}\pm i\underline{\underline{I}}$.
These identities will be used below to obtain the time evolution equation
for the Q-function, and hence the corresponding Fokker-Planck equation. 

\section{Time Evolution \label{sec:Hamiltonian-of-the}}

\subsection{Model Hamiltonian }

To obtain a formalism for the time evolution of the Majorana Q-function
with interacting fermions, we now consider a general Hamiltonian with
a non-interacting linear term and an interaction term. The interaction
Hamiltonian describes a four-Majorana interaction. Depending on the
parameters this Hamiltonian may correspond to the Majorana Hubbard
model~\citep{Affleck_Rahmani_Pikulin_2017,Rahmani:2019MajRev}, or
to a generic four-fermion quantum field theory using a lattice discretization
in space. 

The Hamiltonian of the model is given by:
\begin{align}
\hat{H} & =\hat{H}_{0}+\hat{H}_{int},\nonumber \\
 & =i\hbar\sum_{i,j}t_{ij}\widehat{\gamma}_{i}\widehat{\gamma}_{j}+\frac{\hbar}{2}\sum_{i,j,k,l}g_{ijkl}\widehat{\gamma}_{i}\widehat{\gamma}_{j}\widehat{\gamma}_{k}\widehat{\gamma}_{l}.\label{eq:Ham4Int}
\end{align}
 Due to the antisymmetry of fermion operator commutators, and with
no loss of generality, we can impose the condition that $t_{ij}$
and $g_{ijkl}$ are elements of second and fourth order antisymmetric
tensors respectively. This implies that $t_{ij}=-t_{ji}$ and
\begin{equation}
{\displaystyle g_{ijkl}={\begin{cases}
+g_{\sigma\left(ijkl\right)} & {\text{if }}\sigma(i,j,k,l){\text{ is an even permutation }}\\
-g_{\sigma\left(ijkl\right)} & {\text{if }}\sigma(i,j,k,l){\text{ is an odd permutation }}\\
\;\;\,0 & {\text{otherwise}.}
\end{cases}}}\label{eq:gijkl}
\end{equation}

Since the Hamiltonian is Hermitian, it follows that the coefficients
$t,g$ are all real. This follows since, from hermiticity and antisymmetry,
\begin{align}
t_{ij} & =-t_{ji}^{*}=t_{ij}^{*}\nonumber \\
g_{ijkl} & =g_{lkji}^{*}=g_{ijkl}^{*}.
\end{align}

\subsection{Dynamical evolution }

The time evolution equation for the density operator is given by:
\begin{eqnarray}
i\hbar\frac{\partial}{\partial t}\hat{\rho} & = & \left[\hat{H},\hat{\rho}\right].\label{eq:TimeEvolutionDM-1-1}
\end{eqnarray}

Therefore, the time evolution equation for the Majorana Q-function
obtained from the definition of the Q-function given in Eq (\ref{eq:MajQf})
is:
\begin{equation}
\frac{dQ\left(\underline{\underline{x}}\right)}{dt}=\frac{1}{i\hbar}\mathrm{Tr}\left[\left[\hat{H},\hat{\rho}\right]\hat{\Lambda}^{N}\left(\underline{\underline{x}}\right)\right].\label{eq:TE_MQf-1}
\end{equation}
Using the cyclic identities of the trace, we get:
\begin{equation}
\frac{dQ\left(\underline{\underline{x}}\right)}{dt}=\frac{1}{i\hbar}\mathrm{Tr}\left[\hat{\Lambda}^{N}\hat{H}\hat{\rho}-\hat{H}\hat{\Lambda}^{N}\hat{\rho}\right]\,.\label{eq:TEQ}
\end{equation}
From the Hamiltonian given in Eq. (\ref{eq:Ham4Int}), the time evolution
equation in Eq. (\ref{eq:TEQ}) can be written as:
\begin{eqnarray}
\frac{dQ\left(\underline{\underline{x}}\right)}{dt} & = & \frac{S^{2}}{{\cal N}}\mathrm{Tr}\left\{ t_{ij}\hat{\rho}\left[\hat{\Lambda}\left(\underline{\underline{x}}\right)\widehat{\gamma}_{i}\widehat{\gamma}_{j}-\widehat{\gamma}_{i}\widehat{\gamma}_{j}\hat{\Lambda}\left(\underline{\underline{x}}\right)\right]\right.\nonumber \\
 & + & \frac{\boldsymbol{g}_{\boldsymbol{i}}}{2i}\hat{\rho}\left.\left[\hat{\Lambda}\left(\underline{\underline{x}}\right)\widehat{\gamma}_{i}\widehat{\gamma}_{j}\widehat{\gamma}_{k}\widehat{\gamma}_{l}-\widehat{\gamma}_{i}\widehat{\gamma}_{j}\widehat{\gamma}_{k}\widehat{\gamma}_{l}\hat{\Lambda}\left(\underline{\underline{x}}\right)\right]\right\} .\nonumber \\
\label{eq:TeQCH}
\end{eqnarray}
Here we have defined $g_{ijkl}=\boldsymbol{g}_{\boldsymbol{i}}$,
and $\boldsymbol{i}=\left(i,j,k,l\right)$. We follow the Einstein
summation convention and thereby avoid summation signs throughout
the paper. Repeated indices $i,j,\alpha,\beta,\mu,\nu$ are summed
over $1,\ldots2M$. Using the differential identities of Section \ref{subsec:Majorana-differential-identities}
and the procedure in the Appendix \ref{sec:Appendyx_Dynamical-Ev_Qf},
we obtain:
\begin{eqnarray}
\frac{dQ\left(\underline{\underline{x}}\right)}{dt} & = & -it_{ij}\left(X_{ij}^{\alpha\beta}-X_{ij}^{\alpha\beta*}\right)\partial_{\alpha\beta}Q\label{eq:DQdtX}\\
 & + & \frac{i}{2}\boldsymbol{g}_{\boldsymbol{i}}\left[\left(X_{ij}^{\alpha\beta}X_{kl}^{\mu\nu}-X_{kl}^{\alpha\beta*}X_{ij}^{\mu\nu*}\right)\partial_{\alpha\beta}\partial_{\mu\nu}Q\right.\nonumber \\
 & + & \left(x_{ij}^{+}\left(X_{kl}^{\alpha\beta}-X_{kl}^{\alpha\beta*}\right)+\left(X_{ij}^{\alpha\beta}-X_{ij}^{\alpha\beta*}\right)x_{kl}^{+}\right.\nonumber \\
 & + & \left.\left.X_{ij}^{\mu\nu}\left(\partial_{\mu\nu}X_{kl}^{\alpha\beta}\right)-\left(\partial_{\mu\nu}X_{ij}^{\alpha\beta*}\right)X_{kl}^{\mu\nu*}\right)\partial_{\alpha\beta}Q\right].\nonumber 
\end{eqnarray}

Here we have used the definitions given in Eqs. (\ref{eq:x-x+})
as well as the following properties: 
\begin{align}
X_{\alpha\beta}^{ij*} & =X_{ij}^{\alpha\beta}\nonumber \\
X_{\alpha\beta}^{ji*} & =X_{ij}^{\beta\alpha}\nonumber \\
X_{ij}^{\alpha\beta} & =X_{ji}^{\beta\alpha}.
\end{align}
Since $X_{ij}^{\alpha\beta}-X_{ij}^{\alpha\beta*}=2i\Im X_{ij}^{\alpha\beta}$,
(see Appendix \ref{sec:Appendyx_Dynamical-Ev_Qf}), we can rewrite
the time evolution equation as:
\begin{eqnarray}
\frac{dQ\left(\underline{\underline{x}}\right)}{dt} & = & \left[\frac{i\boldsymbol{g}_{\boldsymbol{i}}}{2}\left(X_{ij}^{\alpha\beta}X_{kl}^{\mu\nu}-X_{kl}^{\alpha\beta*}X_{ij}^{\mu\nu*}\right)\partial_{\mu\nu}\right.\nonumber \\
 & + & \left.2\Im X_{ij}^{\alpha\beta}\left(t_{ij}-3\boldsymbol{g}_{\boldsymbol{i}}x_{kl}^{+}\right)\right]\partial_{\alpha\beta}Q.\label{eq:dQdt}
\end{eqnarray}

We note that Eq. (\ref{eq:dQdt}) sums over all possible values for
the indices marked by Greek labels. Following the procedure described
in Appendix \ref{sec:AppendixSimplificationTEvQIndVar}, we arrive
at the time evolution equation for the Majorana Q-function in terms
of implicit Einstein summation over only independent phase-space variables,
whose indices are denoted $\bm{\alpha}=\left(\alpha,\beta\right)$
with $\alpha<\beta$. These are regarded as a vector, 
\begin{align}
x^{\bm{\alpha}} & \equiv x_{\alpha\beta},\nonumber \\
\partial_{\bm{\alpha}} & \equiv\frac{\partial}{\partial x_{\bm{\alpha}}}.
\end{align}

After making this restriction, and using the antisymmetry of $x_{\alpha\beta}$,
we find that:
\begin{eqnarray}
\frac{dQ\left(\underline{\underline{x}}\right)}{dt} & = & 4\left[\Im X_{ij}^{\bm{\alpha}}\left(t_{ij}-3\boldsymbol{g}_{\boldsymbol{i}}x_{kl}^{+}\right)\partial_{\bm{\alpha}}Q\right.\nonumber \\
 &  & \left.-2\boldsymbol{g}_{\boldsymbol{i}}\left(\Re X_{ij}^{\bm{\alpha}}\right)\left(\Im X_{kl}^{\boldsymbol{\mu}}\right)\partial_{\bm{\alpha}}\partial_{\bm{\mu}}Q\right].\label{eq:dQdtIndv}
\end{eqnarray}
Bold repeated indices $\bm{\alpha},\bm{\mu}$ are summed over independent
variables with $\alpha<\beta$ and $\mu<\nu$. Here the implicit Einstein
summation corresponds to $\sum_{\bm{\alpha}}$ and $\sum_{\bm{\mu}}$
which denote $\sum_{\alpha<\beta}$ and $\sum_{\nu<\mu}$ respectively.
The partial differential equation above can therefore be written in
the form:
\begin{equation}
\frac{dQ\left(\underline{\underline{x}}\right)}{dt}=\left[-\bar{A}^{\bm{\alpha}}+\frac{1}{2}D^{\bm{\alpha}\bm{\mu}}\partial_{\bm{\mu}}\right]\partial_{\bm{\alpha}}Q,\label{eq:PDEQ}
\end{equation}
where
\begin{eqnarray}
D^{\bm{\alpha}\bm{\mu}} & = & -16\boldsymbol{g}_{\boldsymbol{i}}\Re X_{ij}^{\bm{\alpha}}\Im X_{kl}^{\bm{\mu}},\nonumber \\
\bar{A}^{\bm{\alpha}} & = & 4\Im X_{ij}^{\bm{\alpha}}\left(3\boldsymbol{g}_{\boldsymbol{i}}x_{kl}^{+}-t_{ij}\right).\label{eq:driftfirst}
\end{eqnarray}
In this equation, we have denoted the first order coefficients as
$\bar{A}^{\bm{\alpha}}$. In the next section we see that these terms
are related to the diffusion and drift terms of a generalized Fokker-Planck
equation, which has an explicitly probability conserving form. 

\section{Generalized Fokker-Planck equation\label{sec:The-Fokker-Planck}}

In this section we will express the partial differential equation
given in Eq. (\ref{eq:PDEQ}) in the form of a generalized Fokker-Planck
equation. For this purpose, we use the product rule for the second
order derivative, which allows us to write Eq. (\ref{eq:PDEQ}) as,
\begin{eqnarray}
\frac{dQ}{dt} & = & \frac{1}{2}\left[\partial_{\bm{\bm{\alpha}}}\partial_{\bm{\bm{\mu}}}\left(D^{\bm{\alpha}\bm{\mu}}Q\right)-\left(\partial_{\bm{\bm{\alpha}}}\partial_{\bm{\bm{\mu}}}D^{\bm{\alpha}\bm{\mu}}\right)Q\right]\nonumber \\
 &  & -\left(\bar{A}^{\bm{\alpha}}+\left(\partial_{\bm{\mu}}D^{\bm{\alpha}\bm{\mu}}\right)\right)\partial_{\bm{\alpha}}Q,\label{eq:probcon}
\end{eqnarray}
where we use the result that,
\begin{equation}
\partial_{\bm{\alpha}}Q\partial_{\bm{\mu}}D^{\bm{\alpha}\bm{\mu}}+\partial_{\bm{\mu}}Q\partial_{\bm{\alpha}}D^{\bm{\alpha}\bm{\mu}}=2\partial_{\bm{\alpha}}Q\partial_{\bm{\mu}}D^{\bm{\alpha}\bm{\mu}}.
\end{equation}

We have exchanged the dummy indices $\bm{\alpha}\longleftrightarrow\bm{\mu}$
 in the second term of the left hand side of the above equation to
get the desired result. On defining:
\begin{eqnarray}
A^{\bm{\alpha}} & \equiv & \bar{A}^{\bm{\alpha}}+\partial_{\bm{\mu}}D^{\bm{\alpha}\bm{\mu}},\label{eq:DriftTerm}
\end{eqnarray}
we obtain
\begin{eqnarray}
\frac{dQ}{dt} & = & \frac{1}{2}\left(\partial_{\bm{\alpha}}\partial_{\bm{\mu}}D^{\bm{\alpha}\bm{\mu}}-\left[\partial_{\bm{\alpha}}\partial_{\bm{\mu}}D^{\bm{\alpha}\bm{\mu}}\right]\right)Q\nonumber \\
 & - & \left(\partial_{\bm{\alpha}}A^{\bm{\alpha}}-\left[\partial_{\bm{\alpha}}A^{\bm{\alpha}}\right]\right)Q,\label{eq:ProbFin}
\end{eqnarray}
where $A^{\bm{\alpha}}$ is the drift term, while $D^{\bm{\alpha}\bm{\mu}}$
is the diffusion term. Since $\partial_{\bm{\alpha}}\partial_{\bm{\mu}}D^{\bm{\alpha}\bm{\mu}}=0=\partial_{\bm{\alpha}}A^{\bm{\alpha}},$
where the proof is given in Appendix \ref{sec:Deri}, the generalized
Fokker Planck equation for the Majorana Q-function simplifies. It
has an explicitly probability conserving, but not positive-definite,
form \citep{Risken_FP,Carmichael:1999_Book,Drummond:2014BookQT}:
\begin{eqnarray}
\frac{dQ}{dt} & = & \partial_{\bm{\alpha}}\left[-A^{\bm{\alpha}}+\frac{1}{2}\partial_{\bm{\bm{\mu}}}D^{\bm{\alpha}\bm{\mu}}\right]Q.\label{eq:FPE}
\end{eqnarray}

\subsection{Traceless diffusion}

We now investigate whether the diffusion matrix $D^{\bm{\alpha}\bm{\mu}}$
is traceless, as is the case for other Q-function generalized Fokker-Planck
equations. Such equations implement a diffusive 'baker map' transformation,
in which the time-evolution results in mixing, but without the diffusive
growth in entropy of a traditional Fokker-Planck equation \citep{hasegawa1992unitarity,altland2012quantum}.
We will first show that the diffusion term given above includes both
positive and negative-definite forms. 

This term can be written explicitly as:
\begin{eqnarray}
D^{\bm{\alpha}\bm{\mu}} & = & -16\boldsymbol{g}_{\boldsymbol{i}}\Re X_{ij}^{\bm{\alpha}}\Im X_{kl}^{\bm{\mu}}\nonumber \\
 & = & -8\boldsymbol{g}_{\boldsymbol{i}}\left(\Re X_{ij}^{\bm{\alpha}}\Im X_{kl}^{\bm{\mu}}+\Im X_{ij}^{\bm{\alpha}}\Re X_{kl}^{\bm{\mu}}\right).\qquad\label{eq:diffRIS}
\end{eqnarray}
We define the following matrices, that depend only of pairs of indices
$(\alpha,\beta)$ or $(\mu,\nu)$ as:
\begin{eqnarray}
B_{\left(\pm\right)\boldsymbol{i}}^{\bm{\alpha}} & = & \Re X_{ij}^{\bm{\alpha}}\pm\Im X_{kl}^{\bm{\alpha}},\nonumber \\
D_{\left(\pm\right)\boldsymbol{i}}^{\bm{\alpha}\bm{\mu}} & \equiv & B_{\left(\pm\right)\boldsymbol{i}}^{\bm{\alpha}}B_{\left(\pm\right)\boldsymbol{i}}^{\bm{\mu}},\label{eq:Dpmabmn}
\end{eqnarray}
so that the diffusion term is: 
\begin{eqnarray}
D^{\bm{\alpha}\bm{\mu}} & = & 4\boldsymbol{g}_{\boldsymbol{i}}\left(B_{\left(-\right)\boldsymbol{i}}^{\bm{\alpha}}B_{\left(-\right)\boldsymbol{i}}^{\bm{\mu}}-B_{\left(+\right)\boldsymbol{i}}^{\bm{\alpha}}B_{\left(+\right)\boldsymbol{i}}^{\bm{\mu}}\right).
\end{eqnarray}
The matrix is symmetric, and for each set of four indices $\boldsymbol{i}$,
it is expressed as a difference of two real terms, each one being
an outer product of identical vectors. Since any matrix of the form
$D_{\bm{i}}^{\bm{\alpha}\bm{\mu}}=B^{\bm{\alpha}}B^{\bm{\mu}}$ is
positive definite, it follows that each of these matrix product terms
is individually positive definite. Therefore, it is explicitly shown
that the Majorana Q-function diffusion matrix can always be expressed
as the sum of multiple positive-definite and negative-definite terms. 

The diffusion matrix of a standard Fokker Planck equation is symmetric
and positive definite \citep{Risken_FP,Gardiner_Book_SDE}. In the
case of the Q-function this is not the case. Non-positive diffusion
matrices for the generalized Fokker-Planck equations of Q-functions
have been investigated for bosonic and spin systems \citep{Zambrini2003,Altland_PRL2012_Qchaos,Trimborn_PRA79_BH_CS_LieG,Milburn:2002Q}.
This can be interpreted as a forward-backward stochastic process \citep{PM2020_RetrocausalQField,drummond2021time},
i.e. a diffusion that takes place in the forward and backward directions
of time simultaneously.

In terms of the phase-space variables $\underline{\underline{x}}$,
the diffusion matrix is given explicitly as:
\begin{eqnarray}
D^{\bm{\alpha}\bm{\mu}} & = & -4\boldsymbol{g}_{\boldsymbol{i}}\left[\left(x_{i\alpha}x_{\beta j}-x_{i\alpha}\delta_{\beta j}+\delta_{i\alpha}x_{\beta j}+\delta_{i\alpha}\delta_{\beta j}\right)\right.\nonumber \\
 & \times & \left(x_{k\mu}x_{\nu l}-x_{k\mu}\delta_{\nu l}+\delta_{k\mu}x_{\nu l}+\delta_{k\mu}\delta_{\nu l}\right)\nonumber \\
 & - & \left(x_{i\alpha}x_{\beta j}+x_{i\alpha}\delta_{\beta j}-\delta_{i\alpha}x_{\beta j}+\delta_{i\alpha}\delta_{\beta j}\right)\nonumber \\
 & \times & \left.\left(x_{k\mu}x_{\nu l}+x_{k\mu}\delta_{\nu l}-\delta_{k\mu}x_{\nu l}+\delta_{k\mu}\delta_{\nu l}\right)\right].\label{eq:DiffMat}
\end{eqnarray}

We have shown that the diffusion matrix of the Fokker-Planck equation
for the Majorana Q-function can be expressed as a sum of a positive-
and negative-definite terms. We will now show that the diffusion matrix
is completely traceless. In order to prove this, we consider the form
of the diffusion equation given in Eq. (\ref{eq:diffRIS}), which
can also be expressed as:
\begin{eqnarray}
D^{\bm{\alpha}\bm{\mu}} & = & -8\boldsymbol{g}_{\boldsymbol{i}}\left(\Re X_{ij}^{\bm{\alpha}}\Im X_{kl}^{\bm{\mu}}+\Im X_{kl}^{\bm{\alpha}}\Re X_{ij}^{\bm{\mu}}\right)\nonumber \\
 & = & -4\boldsymbol{g}_{\boldsymbol{i}}\Im\left(X_{ij}^{\bm{\alpha}}X_{kl}^{\bm{\mu}}\right).
\end{eqnarray}

To prove the traceless property, it is necessary to show that $\sum_{\bm{\alpha}}D^{\bm{\alpha}\bm{\alpha}}=0$.
We will show this through a proof that every diagonal element of this
matrix in this basis is zero. Each diagonal element has the form:
\begin{align}
D^{\bm{\alpha}\bm{\alpha}} & =-4\boldsymbol{g}_{\boldsymbol{i}}\Im\left(x_{i\alpha}^{+}x_{\beta j}^{-}x_{k\alpha}^{+}x_{\beta l}^{-}\right)\nonumber \\
 & =-4\boldsymbol{g}_{\boldsymbol{i}}\Im\left[\left(x_{i\alpha}+i\delta_{i\alpha}\right)\left(x_{\beta j}-i\delta_{\beta j}\right)\right.\nonumber \\
 & \,\,\,\,\,\,\,\,\,\,\,\,\times\left.\left(x_{k\alpha}+i\delta_{k\alpha}\right)\left(x_{\beta l}-i\delta_{\beta l}\right)\right].
\end{align}
After expanding, we find that the imaginary terms are either cubic
or linear in $\bm{x}$, so that:
\begin{eqnarray}
D^{\bm{\alpha}\bm{\alpha}} & = & 4\boldsymbol{g}_{\boldsymbol{i}}\left[x_{i\alpha}x_{lj}x_{k\alpha}+x_{i\alpha}x_{k\alpha}x_{jl}-x_{ik}x_{\beta j}x_{\beta l}\right.\nonumber \\
 &  & -x_{\beta j}x_{ki}x_{\beta l}-x_{\beta j}\delta_{i\alpha}\delta_{k\alpha}\delta_{\beta l}+x_{i\alpha}\delta_{\beta j}\delta_{k\alpha}\delta_{\beta l}\nonumber \\
 &  & \left.+\delta_{i\alpha}\delta_{\beta j}x_{k\alpha}\delta_{\beta l}-\delta_{i\alpha}\delta_{\beta j}\delta_{k\alpha}x_{\beta l}\right].
\end{eqnarray}

Inspecting these terms, we see that all the terms have similar behavior,
and in each case:
\begin{enumerate}
\item Cubic terms like $\boldsymbol{g}_{\boldsymbol{i}}x_{i\alpha}x_{lj}x_{k\alpha}$
cancel similar terms in the sum with $i$ and $k$ swapped, since
this is an odd permutation which changes the sign of $\boldsymbol{g}_{\boldsymbol{i}}$.
These terms also cancel since $x_{lj}=-x_{jl}.$
\item Linear terms like $\boldsymbol{g}_{\boldsymbol{i}}x_{i\alpha}\delta_{\beta j}\delta_{k\alpha}\delta_{\beta l}$
vanish, since $\boldsymbol{g}_{\boldsymbol{i}}=0$ if $l=j$.
\end{enumerate}
In summary, the diffusion matrix in the $x$ variables has the property
that all the diagonal elements are zero, and consequently it is traceless.
Following the discussion of such traceless equations given elsewhere
\citep{drummond2021time}, it is always possible to make an orthogonal
transformation which diagonalizes the diffusion and leaves the trace
invariant, so that it obeys a forwards-backwards stochastic equation.

\section{Drift term and phase-space domain \label{sec:DriftTermPSD} }

The phase-space variables of the Majorana Q-function are real antisymmetric
matrices, which define a bounded homogeneous phase-space \citep{FermiQ,Ria:2018_Majorana,riaentropy}
with $M(2M-1)$ dimensions \citep{Cartan:1935}. The integration domain,
including the boundary is given by 
\begin{equation}
\underline{\underline{I}}+\underline{\underline{x}}^{2}\geq0.
\end{equation}
 Physical states are characterized by the above condition, which corresponds
to Hermitian, positive density matrices. Gaussian pure fermionic states
are restricted to the surface of the homogeneous space. It is possible
that the generalized Fokker-Planck solutions of Hamiltonian evolution,
even though non-Gaussian, may also be confined to the surface of the
homogeneous space. In this section, we verify this conjecture for
the drift term of the generalized Fokker-Planck equation. 

The condition that defines the the surface of the real subspace of
the complex homogeneous space is given by:
\begin{equation}
\underline{\underline{I}}+\underline{\underline{x}}^{2}=0,
\end{equation}
which can also be written in the form:
\begin{equation}
x_{\alpha\eta}x_{\eta\beta}=-\delta_{\alpha\beta}.\label{eq:CSPS}
\end{equation}
On differentiating the above equation we obtain:
\begin{equation}
\frac{\partial x_{\alpha\eta}}{\partial t}x_{\eta\beta}+x_{\alpha\eta}\frac{\partial x_{\eta\beta}}{\partial t}=-\frac{\partial}{\partial t}\delta_{\alpha\beta}=0.\label{eq:DifC}
\end{equation}
We now wish to relate this condition with the drift term, $A^{\left(\alpha\eta\right)}$
and the surface of the homogeneous space. On considering the generic
drift equation, 
\begin{equation}
\frac{\partial x_{\bm{\alpha}}}{\partial t}\equiv A^{\bm{\alpha}},
\end{equation}
Eq. (\ref{eq:DifC}) can be written as:
\begin{equation}
A^{\left(\alpha\eta\right)}x_{\eta\beta}+x_{\alpha\eta}A^{\left(\eta\beta\right)}=0.\label{eq:CA}
\end{equation}
Using Eq. (\ref{eq:DriftTerm}) as well as Eq. (\ref{eq:driftfirst})
the expression for the drift term is given by:
\begin{equation}
A^{\left(\alpha\beta\right)}=-\frac{4\boldsymbol{g}_{\boldsymbol{i}}}{\hbar}\Im X_{ij}^{\alpha\beta}\left(6x_{kl}^{+}-\frac{t_{ij}}{\boldsymbol{g}_{\boldsymbol{i}}}+8\left(3-2M\right)x_{kl}\right).\label{eq:DriftTermEE}
\end{equation}

On substituting this expression in the left hand side of Eq. (\ref{eq:CA})
we obtain:
\begin{eqnarray}
 &  & x_{\alpha\eta}A^{\eta\beta}+A^{\alpha\eta}x_{\eta\beta}=\nonumber \\
 &  & -\frac{4\boldsymbol{g}_{\boldsymbol{i}}}{\hbar}\left[\Im X_{ij}^{\eta\beta}\left(-x_{\alpha\eta}\frac{t_{ij}}{\boldsymbol{g}_{\boldsymbol{i}}}+2\left(15-8M\right)x_{\alpha\eta}x_{kl}^{+}\right)\right.\nonumber \\
 &  & \left.+\Im X_{ij}^{\alpha\eta}\left(-\frac{t_{ij}}{\boldsymbol{g}_{\boldsymbol{i}}}x_{\eta\beta}+2\left(15-8M\right)x_{kl}^{+}x_{\eta\beta}\right)\right]\nonumber \\
 &  & =\frac{4\boldsymbol{g}_{\boldsymbol{i}}}{\hbar}\left[2\left(15-8M\right)\left(-\delta_{\beta j}x_{i\eta}x_{\eta\alpha}+\delta_{i\alpha}x_{j\eta}x_{\eta\beta}\right)x_{kl}^{+}\right.\nonumber \\
 &  & \left.-\frac{t_{ij}}{\boldsymbol{g}_{\boldsymbol{i}}}\left(-\delta_{\beta j}x_{i\eta}x_{\eta\alpha}+\delta_{i\alpha}x_{j\eta}x_{\eta\beta}\right)\right].\label{eq:xA}
\end{eqnarray}
Here we have used that,
\begin{eqnarray*}
\Im X_{ij}^{\eta\beta}x_{\alpha\eta} & = & -\left(x_{i\eta}\delta_{\beta j}x_{\alpha\eta}-x_{\beta j}x_{\alpha i}\right).
\end{eqnarray*}

We notice that the expression of Eq. (\ref{eq:xA}) contains terms
of the form $x_{i\eta}x_{\eta\alpha}$. If we consider the condition
for the pure states on the boundary of the homogeneous space, given
in Eq. (\ref{eq:CSPS}) we get:
\begin{eqnarray}
 &  & x_{\alpha\eta}A^{\eta\beta}+A^{\alpha\eta}x_{\eta\beta}\nonumber \\
 &  & =\frac{4\boldsymbol{g}_{\boldsymbol{i}}}{\hbar}\left[2\left(15-8M\right)\left(\delta_{\beta j}\delta_{i\alpha}-\delta_{i\alpha}\delta_{j\beta}\right)x_{kl}^{+}\right.\nonumber \\
 &  & \left.-\frac{t_{ij}}{\boldsymbol{g}_{\boldsymbol{i}}}\left(\delta_{\beta j}\delta_{i\alpha}-\delta_{i\alpha}\delta_{j\beta}\right)\right]=0.
\end{eqnarray}

In summary, the drift term maintains the `surface' condition that
corresponds to a Gaussian pure state. It is a tangent vector in the
space.

\section{Summary\label{sec:Summary}}

We have considered a completely general four-fermion interaction Hamiltonian
that contains four Majorana operators. For this model, we have derived
the time evolution equation for the Majorana Q-function phase-space
representation. This type of interaction Hamiltonian can be used to
describe the Majorana-Hubbard and Fermi-Hubbard models, as well as
more general Hamiltonians in quantum field theory. In order to perform
the calculations we have used the symmetry properties of the Hamiltonian.
We have derived a generalized Fokker-Planck type equation, whose diffusion
term is not positive definite. Instead we show it has a zero trace:
the diffusion term can be expressed as a sum of positive definite
and negative definite terms. This is consistent with a forward-backward
stochastic evolution, as found previously for the evolution of bosonic
and spin Q-functions. Such evolution has a probabilistic action and
path integral \citep{drummond2021time}, compatible with an ontological
interpretation \citep{PM2020_RetrocausalQField} as an objective field
without requiring observers \citep{bong2020strong}. 
\begin{acknowledgments}
This research was supported by Australian Research Council Discovery
Project Grant DP190101480.
\end{acknowledgments}

\appendix

\section{Dynamics of Majorana Q-functions\label{sec:Appendyx_Dynamical-Ev_Qf}}

In this section we give details of the calculations used to obtain
the time evolution of the Majorana Q-function given in Eq. (\ref{eq:DQdtX}).
All repeated indices are summed over their full range of definition,
where $i,j,k,l=1,\ldots M$ and $\mu,\nu,\alpha,\beta=1,\ldots2M$
. The bold notation $\bm{\mu}=\left(\mu,\nu\right),\bm{\alpha}=\left(\alpha,\beta\right)$
indicates ordered indices summed with $\mu<\nu,\alpha<\beta=1,\ldots2M$.

First, consider the time evolution equation of the Majorana Q-function,
Eq. (\ref{eq:TeQCH}), which is: 
\begin{eqnarray}
\frac{dQ\left(\underline{\underline{x}}\right)}{dt} & = & \frac{S^{2}}{{\cal N}}\mathrm{Tr}\left\{ t_{ij}\hat{\rho}\left[\hat{\Lambda}\left(\underline{\underline{x}}\right)\widehat{\gamma}_{i}\widehat{\gamma}_{j}-\widehat{\gamma}_{i}\widehat{\gamma}_{j}\hat{\Lambda}\left(\underline{\underline{x}}\right)\right]\right.\\
 & + & \frac{\boldsymbol{g}_{\boldsymbol{i}}}{2i}\hat{\rho}\left.\left[\hat{\Lambda}\left(\underline{\underline{x}}\right)\widehat{\gamma}_{i}\widehat{\gamma}_{j}\widehat{\gamma}_{k}\widehat{\gamma}_{l}-\widehat{\gamma}_{i}\widehat{\gamma}_{j}\widehat{\gamma}_{k}\widehat{\gamma}_{l}\hat{\Lambda}\left(\underline{\underline{x}}\right)\right]\right\} .\nonumber 
\end{eqnarray}

In calculating the time evolution equation, products of the form $\widehat{\gamma}_{i}\widehat{\gamma}_{j}$
and $\widehat{\gamma}_{i}\widehat{\gamma}_{j}\widehat{\gamma}_{k}\widehat{\gamma}_{l}$
must be evaluated. The differential identities given in Eqs. (\ref{eq:UnorderedDifId})
and (\ref{eq:UnorderedDifId-1}) are used to obtain:
\begin{eqnarray}
\widehat{\gamma}_{i}\widehat{\gamma}_{j}\widehat{\gamma}_{k}\widehat{\gamma}_{l}\hat{\Lambda} & = & \left(-X_{ij}^{\mu\nu*}\partial_{\alpha\beta}\partial_{\mu\nu}\hat{\Lambda}+\partial_{\mu\nu}\hat{\Lambda}\partial_{\beta\alpha}X_{ij}^{\mu\nu*}\right)X_{kl}^{\alpha\beta*}\nonumber \\
 & + & \left(-\partial_{\alpha\beta}\hat{\Lambda}x_{ij}^{+}-\hat{\Lambda}\partial_{\alpha\beta}\hat{\Lambda}x_{ij}^{+}\right)X_{kl}^{\alpha\beta*}\nonumber \\
 &  & -X_{ij}^{\mu\nu*}x_{kl}^{+}\partial_{\mu\nu}\hat{\Lambda}+\hat{\Lambda}X_{ik}^{jl}.\label{eq:DIQOR}
\end{eqnarray}
\begin{eqnarray}
\hat{\Lambda}\widehat{\gamma}_{i}\widehat{\gamma}_{j}\widehat{\gamma}_{k}\widehat{\gamma}_{l} & = & X_{ij}^{\alpha\beta}\left(-X_{kl}^{\mu\nu}\partial_{\alpha\beta}\partial_{\mu\nu}\hat{\Lambda}-\partial_{\mu\nu}\hat{\Lambda}\partial_{\alpha\beta}X_{kl}^{\mu\nu}\right)\nonumber \\
 & + & X_{ij}^{\alpha\beta}\left(-\left(\partial_{\alpha\beta}\hat{\Lambda}\right)x_{kl}^{+}-\hat{\Lambda}\partial_{\alpha\beta}x_{kl}^{+}\right)\nonumber \\
 &  & -x_{ij}^{+}X_{kl}^{\mu\nu}\partial_{\mu\nu}\hat{\Lambda}+\hat{\Lambda}X_{ik}^{jl}.\label{eq:DIQOL}
\end{eqnarray}
 Here, the definitions given in Eq. (\ref{eq:x-x+}) are utilized
so that $X_{ij}^{\alpha\beta}=x_{i\alpha}^{+}x_{\beta j}^{-},$ and
$X_{ij}^{\alpha\beta*}=x_{i\alpha}^{-}x_{\beta j}^{+}.$ The derivative
of the Gaussian basis and the derivative of the Gaussian operator
are related through the chain rule as:
\begin{equation}
\frac{1}{{\cal N}}S\left(\left[\underline{\underline{x}}\right]^{2}\right)\frac{d\hat{\Lambda}\left(\underline{\underline{x}}\right)}{d\underline{\underline{x}}}=\frac{d\hat{\Lambda}^{N}\left(\underline{\underline{x}}\right)}{d\underline{\underline{x}}}-\frac{d\ln S\left(\left[\underline{\underline{x}}\right]^{2}\right)}{d\underline{\underline{x}}}\hat{\Lambda}^{N}\left(\underline{\underline{x}}\right).
\end{equation}
We now take the limit $S\rightarrow1$, so that 
\begin{equation}
\frac{1}{{\cal N}}S\left(\left[\underline{\underline{x}}\right]^{2}\right)\frac{d\hat{\Lambda}\left(\underline{\underline{x}}\right)}{d\underline{\underline{x}}}=\frac{d\hat{\Lambda}^{N}\left(\underline{\underline{x}}\right)}{d\underline{\underline{x}}},
\end{equation}
and $\hat{\Lambda}^{N}\left(\underline{\underline{x}}\right)=\frac{1}{{\cal N}}\hat{\Lambda}\left(\underline{\underline{x}}\right).$
Using the differential identities given in Eqs. (\ref{eq:UnorderedDifId}),
(\ref{eq:UnorderedDifId-1}), (\ref{eq:DIQOR}) and (\ref{eq:DIQOL})
gives:
\begin{eqnarray}
\frac{dQ\left(\underline{\underline{x}}\right)}{dt} & = & 2t_{ij}\Im X_{ij}^{\alpha\beta}\partial_{\alpha\beta}Q+\nonumber \\
 &  & \frac{i}{2}\boldsymbol{g}_{\boldsymbol{i}}\left[\left(x_{kl}^{+}X_{ij}^{\alpha\beta}-x_{ij}^{+}X_{kl}^{\alpha\beta*}\right)\partial_{\alpha\beta}\right.\nonumber \\
 & + & \left(X_{ij}^{\alpha\beta}\partial_{\alpha\beta}X_{kl}^{\mu\nu}-\partial_{\alpha\beta}X_{ij}^{\mu\nu*}X_{kl}^{\alpha\beta*}-X_{ij}^{\mu\nu*}x_{kl}^{+}\right.\nonumber \\
 & + & \left.x_{ij}^{+}X_{kl}^{\mu\nu}\right)\partial_{\mu\nu}Q+\left(X_{ij}^{\alpha\beta}\partial_{\alpha\beta}x_{kl}^{+}-X_{kl}^{\alpha\beta*}\partial_{\alpha\beta}x_{ij}^{+}\right)\nonumber \\
 &  & +.\left.\left(X_{ij}^{\alpha\beta}X_{kl}^{\mu\nu}-X_{kl}^{\alpha\beta*}X_{ij}^{\mu\nu*}\right)\partial_{\alpha\beta}\partial_{\mu\nu}\right]Q.
\end{eqnarray}
Here we have defined $X_{ij}^{\alpha\beta}-X_{ij}^{\alpha\beta*}=2i\Im X_{ij}^{\alpha\beta}.$
Next, the expressions given in Eq. (\ref{eq:Der}) lead to:
\begin{eqnarray*}
\partial_{\alpha\beta}x_{kl}^{+}X_{ij}^{\alpha\beta}-\partial_{\alpha\beta}x_{ij}^{+}X_{kl}^{\alpha\beta*} & = & 0.
\end{eqnarray*}
 Since the indices $\alpha$, $\beta$, $\mu$ and $\nu$ are dummy
indices we can interchange $\alpha\rightarrow\beta$ and $\mu\rightarrow\nu$
in the terms that multiply $\partial_{\mu\nu}Q$. Therefore we obtain:
\begin{eqnarray}
\frac{dQ\left(\underline{\underline{x}}\right)}{dt} & = & 2t_{ij}\Im X_{ij}^{\alpha\beta}\partial_{\alpha\beta}Q+\\
 & + & \frac{i}{2}\boldsymbol{g}_{\boldsymbol{i}}\left[\left(X_{ij}^{\alpha\beta}X_{kl}^{\mu\nu}-X_{kl}^{\alpha\beta*}X_{ij}^{\mu\nu*}\right)\partial_{\alpha\beta}\partial_{\mu\nu}\right.\nonumber \\
 & + & \left(2i\left(\Im X_{kl}^{\alpha\beta}x_{ij}^{+}+\Im X_{ij}^{\alpha\beta}x_{kl}^{+}\right)\right.\nonumber \\
 & + & \left.\left.X_{ij}^{\mu\nu}\left(\partial_{\mu\nu}X_{kl}^{\alpha\beta}\right)-X_{kl}^{\mu\nu*}\left(\partial_{\mu\nu}X_{ij}^{\alpha\beta*}\right)\right)\partial_{\alpha\beta}\right]Q.\nonumber 
\end{eqnarray}

To simplify the results further, we swap $i\longleftrightarrow k$
and $\longleftrightarrow j\longleftrightarrow l$ in the terms $x_{\alpha\beta}\Im X_{kl}^{\alpha\beta}x_{ij}^{+}$
and $g_{ijkl}X_{ij}^{\mu\nu}\partial_{\mu\nu}X_{kl}^{\alpha\beta}$.
There is no sign change in $\boldsymbol{g}_{\boldsymbol{i}}$ since
it is an even permutation, so we obtain:
\begin{eqnarray}
\frac{dQ\left(\underline{\underline{x}}\right)}{dt} & = & 2t_{ij}\Im X_{ij}^{\alpha\beta}\partial_{\alpha\beta}Q+\\
 &  & \frac{i}{2}\boldsymbol{g}_{\boldsymbol{i}}\left[\left(X_{ij}^{\alpha\beta}X_{kl}^{\mu\nu}-X_{kl}^{\alpha\beta*}X_{ij}^{\mu\nu*}\right)\partial_{\alpha\beta}\partial_{\mu\nu}\right.\nonumber \\
 & + & \left(2i\left(\Im X_{ij}^{\alpha\beta}x_{kl}^{+}+\Im X_{ij}^{\alpha\beta}x_{kl}^{+}\right)\right.\nonumber \\
 & + & \left.\left.X_{kl}^{\mu\nu}\left(\partial_{\mu\nu}X_{ij}^{\alpha\beta}\right)-\left(\partial_{\mu\nu}X_{ij}^{\alpha\beta*}\right)X_{kl}^{\mu\nu*}\right)\partial_{\alpha\beta}\right]Q.\nonumber 
\end{eqnarray}
Using the derivative properties described in Section \ref{subsec:usefulderivatives},
as well as the swapping of indices and the permutation properties
of $\boldsymbol{g}_{\boldsymbol{i}}$, we get:
\begin{eqnarray*}
\boldsymbol{g}_{\boldsymbol{i}}\left(X_{kl}^{\mu\nu}\partial_{\mu\nu}X_{ij}^{\alpha\beta}-X_{kl}^{\mu\nu*}\partial_{\mu\nu}X_{ij}^{\alpha\beta*}\right) & = & 8i\boldsymbol{g}_{\boldsymbol{i}}x_{kl}^{+}\Im X_{ij}^{\alpha\beta}.
\end{eqnarray*}
Therefore we finally obtain the simplified form of the time evolution
equation of the Majorana $Q$-function:
\begin{eqnarray}
\frac{dQ\left(\underline{\underline{x}}\right)}{dt} & = & i\left[\frac{\boldsymbol{g}_{\boldsymbol{i}}}{2}\left(X_{ij}^{\alpha\beta}X_{kl}^{\mu\nu}-X_{kl}^{\alpha\beta*}X_{ij}^{\mu\nu*}\right)\partial_{\alpha\beta}\partial_{\mu\nu}\right.\nonumber \\
 & + & \left.2i\Im X_{ij}^{\alpha\beta}\left(3\boldsymbol{g}_{\boldsymbol{i}}x_{kl}^{+}-t_{ij}\right)\partial_{\alpha\beta}\right]Q.\label{eq:dQdtAp}
\end{eqnarray}
This equation corresponds to Eq. (\ref{eq:dQdt}).

\section{Identities with independent variables\label{sec:AppendixSimplificationTEvQIndVar}}

Since the $\underline{\underline{x}}$ matrices are real antisymmetric
matrices, $x_{\alpha\beta}$ and $x_{\beta\alpha}$ are not independent.
However, Eq. (\ref{eq:dQdtAp}) includes all possible values for the
indices labeled by Greek letters. This implies, for example, for the
term involving second order derivatives, that:
\begin{align}
\boldsymbol{g}_{\boldsymbol{i}}X_{ij}^{\alpha\beta}X_{kl}^{\mu\nu}\partial_{\alpha\beta}\partial_{\mu\nu}Q & =\sum_{\alpha<\beta\mu<\nu}\boldsymbol{g}_{\boldsymbol{i}}X_{ij}^{\alpha\beta}X_{kl}^{\mu\nu}\partial_{\alpha\beta}\partial_{\mu\nu}Q\nonumber \\
 & +\sum_{\alpha>\beta\mu>\nu}\boldsymbol{g}_{\boldsymbol{i}}X_{ij}^{\alpha\beta}X_{kl}^{\mu\nu}\partial_{\alpha\beta}\partial_{\mu\nu}Q\nonumber \\
 & +\sum_{\alpha<\beta\mu>\nu}\boldsymbol{g}_{\boldsymbol{i}}X_{ij}^{\alpha\beta}X_{kl}^{\mu\nu}\partial_{\alpha\beta}\partial_{\mu\nu}Q\nonumber \\
 & +\sum_{\alpha>\beta\mu<\nu}\boldsymbol{g}_{\boldsymbol{i}}X_{ij}^{\alpha\beta}X_{kl}^{\mu\nu}\partial_{\alpha\beta}\partial_{\mu\nu}Q.
\end{align}

In order to write Eq. (\ref{eq:dQdt}), only considering independent
variables, the following steps will be used. In the second term of
the above expression one can swap the dummy indices $\alpha\longleftrightarrow\beta$
and $\mu\longleftrightarrow\nu$. We can rewrite the derivative term
by using: $x_{\beta\alpha}=-x_{\alpha\beta}$ and $x_{\nu\mu}=-x_{\mu\nu}$.
Latin indices can be swapped as $i\longrightarrow j$ and $k\longrightarrow l$.
This swapping does not change the sign of $\boldsymbol{g}_{\boldsymbol{i}}$,
since it is an even permutation, so the second term becomes $\sum_{\alpha<\beta\mu<\nu}\boldsymbol{g}_{\boldsymbol{i}}X_{ji}^{\beta\alpha}X_{lk}^{\nu\mu}\partial_{\alpha\beta}\partial_{\mu\nu}Q$.
Since $X_{ij}^{\alpha\beta}=X_{ji}^{\beta\alpha}$, we notice that,
after this swapping, the second term and the first term of the above
expression are the same. 

Following an analogous procedure to that described above, we can perform
the corresponding swapping in the third and fourth term, showing that
all four terms are equivalent. Hence, the above expression is written
using independent variables, as: 
\begin{eqnarray}
{\normalcolor \boldsymbol{g}_{\boldsymbol{i}}X_{ij}^{\alpha\beta}X_{kl}^{\mu\nu}\partial_{\alpha\beta}\partial_{\mu\nu}Q} & \;{\normalcolor =} & \;{\normalcolor 4\sum_{\alpha<\beta\mu<\nu}\boldsymbol{g}_{\boldsymbol{i}}X_{ij}^{\alpha\beta}X_{kl}^{\mu\nu}\partial_{\alpha\beta}\partial_{\mu\nu}Q}\nonumber \\
{\normalcolor } & \;{\normalcolor \equiv} & \;4{\normalcolor \boldsymbol{g}_{\boldsymbol{i}}X_{ij}^{\bm{\alpha}}X_{kl}^{\bm{\mu}}\partial_{\bm{\bm{\alpha}}}\partial_{\bm{\bm{\mu}}}Q.}\label{eq:restsumm}
\end{eqnarray}

\textcolor{black}{In the last line of the above equation we use the
notation that bold repeated indices $\bm{\alpha},\bm{\mu}$ are summed
over independent variables with $\alpha<\beta$ and $\mu<\nu$. }We
will also make a restricted summation of the coefficients of $\partial_{\alpha\beta}Q$,
again in order to only include independent variables. Following an
analogous procedure, this gives an additional overall factor of $2.$
We will explain the procedure for the term that corresponds to the
linear Hamiltonian, which is $2i\Im X_{ij}^{\alpha\beta}t_{ij}\partial_{\alpha\beta}Q$.
In this case,
\begin{eqnarray}
2i\Im X_{ij}^{\alpha\beta}t_{ij}\partial_{\alpha\beta}Q & = & t_{ij}\sum_{\alpha<\beta}2i\Im X_{ij}^{\alpha\beta}\partial_{\alpha\beta}Q\nonumber \\
 & + & t_{ij}\sum_{\beta<\alpha}2i\Im X_{ij}^{\alpha\beta}\partial_{\alpha\beta}Q.
\end{eqnarray}
Next, swapping the Latin indices $i\longleftrightarrow j,$ gives
$-2i\Im X_{ij}^{\alpha\beta}t_{ji}\partial_{\alpha\beta}Q.$ Since
$t_{ij}=-t_{ji,}$ and $X_{ji}^{\beta\alpha}=X_{ij}^{\alpha\beta},$
the second term is identical to the first term of the above expressions.
Thus, in terms of independent variables the term corresponding to
the linear Hamiltonian is:
\begin{equation}
2i\Im\sum_{\alpha,\beta}X_{ij}^{\alpha\beta}t_{ij}\partial_{\alpha\beta}Q=4it_{ij}\Im X_{ij}^{\bm{\alpha}}\partial_{\bm{\bm{\alpha}}}Q.
\end{equation}

Following this method for the other terms, and using implicit Einstein
summation over $\bm{\alpha}$ and $\bm{\mu}$ in terms of only independent
variables with $\alpha<\beta$ and $\mu<\nu$, Eq. (\ref{eq:dQdtAp})
becomes:
\begin{eqnarray}
\frac{dQ\left(\underline{\underline{x}}\right)}{dt} & = & \left[4\Im X_{ij}^{\bm{\bm{\alpha}}}\left(t_{ij}-3\boldsymbol{g}_{\boldsymbol{i}}x_{kl}^{+}\right)\partial_{\bm{\bm{\alpha}}}Q\right.\nonumber \\
 & + & \left.2\boldsymbol{g}_{\boldsymbol{i}}\left(X_{ij}^{\bm{\bm{\alpha}}}X_{kl}^{\bm{\bm{\mu}}}-X_{kl}^{\bm{\bm{\alpha}}*}X_{ij}^{\bm{\bm{\mu}}*}\right)\partial_{\bm{\bm{\alpha}}}\partial_{\bm{\bm{\mu}}}Q\right].\nonumber \\
\label{eq:dQdtSimp}
\end{eqnarray}
This equation corresponds to Eq. (\ref{eq:dQdtIndv}). 

On performing exchanges of indices, the following symmetry properties
were used:
\begin{itemize}
\item Symmetry of $x$ matrices: $x_{\alpha\beta}=-x_{\beta\alpha}$.
\item Symmetry of $\boldsymbol{g}_{\boldsymbol{i}}$: One can always exchange
the Latin indices $i,j,k$ and $l$ while taking into account the
number of permutations between indices.
\end{itemize}
We now simplify the term $\boldsymbol{g}_{\boldsymbol{i}}\left(X_{ij}^{\alpha\beta}X_{kl}^{\mu\nu}-X_{ij}^{\mu\nu*}X_{kl}^{\alpha\beta*}\right)\partial_{\bm{\bm{\alpha}}}\partial_{\bm{\bm{\mu}}}Q.$
To do this, the symmetry properties above are used, together with:
\begin{itemize}
\item Swapping on restricted summations: One can always exchange $\alpha$
with $\mu$ and $\beta$ with $\nu$ simultaneously without any sign
change. The second order derivative of the $Q$-function does not
change, as $\partial_{\bm{\bm{\alpha}}}\partial_{\bm{\bm{\mu}}}Q=\partial_{\bm{\bm{\mu}}}\partial_{\bm{\bm{\alpha}}}Q$.
\end{itemize}
Using these symmetry properties, we obtain $\boldsymbol{g}_{\boldsymbol{i}}X_{kl}^{\alpha\beta*}X_{ij}^{\mu\nu*}=\boldsymbol{g}_{\boldsymbol{i}}\left(X_{ij}^{\alpha\beta}X_{kl}^{\mu\nu}\right)^{*},$
and also that:
\begin{eqnarray}
X_{ij}^{\alpha\beta}X_{kl}^{\mu\nu}-X_{kl}^{\alpha\beta*}X_{ij}^{\mu\nu*} & = & X_{ij}^{\alpha\beta}X_{kl}^{\mu\nu}-\left(X_{ij}^{\alpha\beta}X_{kl}^{\mu\nu}\right)^{*}\nonumber \\
 & = & 2i\Im(X_{ij}^{\alpha\beta}X_{kl}^{\mu\nu})\nonumber \\
 & = & 2i\left(\Re X_{ij}^{\alpha\beta}\Im X_{kl}^{\mu\nu}+\Im X_{ij}^{\alpha\beta}\Re X_{kl}^{\mu\nu}\right).\nonumber \\
\end{eqnarray}
Therefore, we see that: 
\begin{eqnarray}
 &  & 2i\boldsymbol{g}_{\boldsymbol{i}}\left(X_{ij}^{\alpha\beta}X_{kl}^{\mu\nu}-X_{kl}^{\alpha\beta*}X_{ij}^{\mu\nu*}\right)\nonumber \\
 &  & =-4\boldsymbol{g}_{\boldsymbol{i}}\left(\Re X_{ij}^{\alpha\beta}\Im X_{kl}^{\mu\nu}+\Im X_{ij}^{\alpha\beta}\Re X_{kl}^{\mu\nu}\right).
\end{eqnarray}
 We now exchange $i\longleftrightarrow k$ and $j\longleftrightarrow l$,
in the term $\boldsymbol{g}_{\boldsymbol{i}}\Im X_{ij}^{\alpha\beta}\Re X_{kl}^{\mu\nu}$,
there is no sign change in $\boldsymbol{g}_{\boldsymbol{i}},$ since
it is an even permutation, then $\boldsymbol{g}_{\boldsymbol{i}}\Im X_{ij}^{\alpha\beta}\Re X_{kl}^{\mu\nu}=\boldsymbol{g}_{\boldsymbol{i}}\Im X_{kl}^{\alpha\beta}\Re X_{ij}^{\mu\nu}$.
Next we swap the dummy indices $\alpha\longleftrightarrow\mu$ and
$\beta\longleftrightarrow\nu$, obtaining $\Im X_{kl}^{\alpha\beta}\Re X_{ij}^{\mu\nu}=\Im X_{kl}^{\mu\nu}\Re X_{ij}^{\alpha\beta}.$
Therefore we get:
\begin{eqnarray}
2i\boldsymbol{g}_{\boldsymbol{i}}\left(X_{ij}^{\alpha\beta}X_{kl}^{\mu\nu}-X_{kl}^{\alpha\beta*}X_{ij}^{\mu\nu*}\right) & = & -8\boldsymbol{g}_{\boldsymbol{i}}\Re X_{ij}^{\alpha\beta}\Im X_{kl}^{\mu\nu}.\qquad
\end{eqnarray}
In terms of the phase-space variables $x_{i\alpha}$, the term $\Re X_{ij}^{\alpha\beta}\Im X_{kl}^{\mu\nu}$
is:
\begin{eqnarray}
\Re X_{ij}^{\alpha\beta}\Im X_{kl}^{\mu\nu} & = & \left(x_{i\alpha}x_{\beta j}+\delta_{i\alpha}\delta_{\beta j}\right)\left(-x_{k\mu}\delta_{\nu l}+\delta_{k\mu}x_{\nu l}\right).\qquad
\end{eqnarray}

This concludes the derivation of the generalized Fokker-Planck equation
given in the main text.

\section{Derivatives\label{sec:Deri}}

In this section we give details of the calculations of the first order
derivatives of the diffusion and drift term, as well as the second
order derivative of the diffusion term of the Fokker-Planck equation
for the Majorana Q-function. 

\subsection{Derivatives of the diffusion term}

In order to calculate the first order derivative of the diffusion
term, we use its symmetric form, as in Eq. (\ref{eq:diffRIS}), 
as shown below:
\begin{eqnarray}
\partial_{\bm{\mu}}D^{\bm{\alpha}\bm{\mu}} & = & -8\boldsymbol{g}_{\boldsymbol{i}}\partial_{\bm{\mu}}\left[\Re X_{ij}^{\alpha\beta}\Im X_{kl}^{\mu\nu}+\Im X_{kl}^{\alpha\beta}\Re X_{ij}^{\mu\nu}\right]\nonumber \\
 & = & -8\boldsymbol{g}_{\boldsymbol{i}}\left[\partial_{\bm{\mu}}\Re X_{ij}^{\alpha\beta}\Im X_{kl}^{\mu\nu}+\Re X_{ij}^{\alpha\beta}\partial_{\bm{\mu}}\Im X_{kl}^{\mu\nu}+\right.\nonumber \\
 &  & \left.\partial_{\bm{\mu}}\Im X_{kl}^{\alpha\beta}\Re X_{ij}^{\mu\nu}+\Im X_{kl}^{\alpha\beta}\partial_{\bm{\mu}}\Re X_{ij}^{\mu\nu}\right].
\end{eqnarray}
Using the product rule of the derivative as well as the derivatives
of Section \ref{subsec:usefulderivatives}, we get the following results:
\begin{eqnarray}
\Im X_{kl}^{\mu\nu}\partial_{\bm{\mu}}\Re X_{ij}^{\alpha\beta} & = & x_{\beta j}\left(\delta_{l\alpha}x_{ik}-\delta_{k\alpha}x_{il}+x_{k\alpha}\delta_{il}-x_{l\alpha}\delta_{ki}\right)\nonumber \\
 & + & x_{i\alpha}\left(\delta_{kj}x_{l\beta}-x_{k\beta}\delta_{jl}-x_{jk}\delta_{\beta l}+\delta_{k\beta}x_{jl}\right),\nonumber \\
\partial_{\bm{\mu}}\Im X_{kl}^{\mu\nu} & = & -\delta_{kl}\left(1-2M\right)+\delta_{lk}\left(1-2M\right)=0,\nonumber \\
\Re X_{ij}^{\mu\nu}\partial_{\bm{\mu}}\Im X_{kl}^{\alpha\beta} & = & \delta_{\beta l}\left(x_{ik}x_{j\alpha}-x_{i\alpha}x_{jk}-\delta_{ik}\delta_{\alpha j}+\delta_{\alpha i}\delta_{jk}\right)\nonumber \\
 & + & \delta_{k\alpha}\left(x_{il}x_{j\beta}-x_{i\beta}x_{jl}+\delta_{i\beta}\delta_{lj}-\delta_{j\beta}\delta_{li}\right),\nonumber \\
\Im X_{kl}^{\alpha\beta}\partial_{\bm{\mu}}\Re X_{ij}^{\mu\nu} & = & 2x_{ij}\left(1-2M\right)\left(-x_{k\alpha}\delta_{\beta l}+\delta_{k\alpha}x_{\beta l}\right).
\end{eqnarray}
After substituting the above results and on simplifying terms, gives:
\begin{eqnarray}
\partial_{\bm{\mu}}D^{\bm{\alpha}\bm{\mu}} & = & -8\boldsymbol{g}_{\boldsymbol{i}}\left[x_{ik}x_{\beta j}\delta_{l\alpha}+\delta_{\beta l}x_{ik}x_{j\alpha}-2x_{il}x_{\beta j}\delta_{k\alpha}\right.\nonumber \\
 &  & +2x_{kj}x_{i\alpha}\delta_{\beta l}-x_{lj}x_{i\alpha}\delta_{k\beta}+x_{i\beta}x_{lj}\delta_{k\alpha}\nonumber \\
 & + & \left.2x_{ij}\left(1-2M\right)\left(-x_{k\alpha}\delta_{\beta l}+\delta_{k\alpha}x_{\beta l}\right)\right].\label{eq:DerDxmunuNS}
\end{eqnarray}
Here we have also considered that we are considering $i\neq j\neq k\neq l$.
As in previous calculations, we use symmetry properties in order to
simplify the term given in Eq. (\ref{eq:DerDxmunuNS}). We perform
the swapping given in Table \ref{tab:TransfTable}, obtaining:
\begin{eqnarray}
\partial_{\bm{\mu}}D^{\bm{\alpha}\bm{\mu}} & = & 16\left(3-2M\right)\boldsymbol{g}_{\boldsymbol{i}}x_{ij}\left(x_{k\alpha}\delta_{\beta l}-\delta_{k\alpha}x_{\beta l}\right)\nonumber \\
 & = & -16\left(3-2M\right)\boldsymbol{g}_{\boldsymbol{i}}x_{ij}\Im X_{kl}^{\alpha\beta}\nonumber \\
 & = & -16\left(3-2M\right)\boldsymbol{g}_{\boldsymbol{i}}x_{kl}\Im X_{ij}^{\alpha\beta}.\label{eq:DermunuDifTerm}
\end{eqnarray}
In the last line of the above equation we have exchanged the indices
$i\longleftrightarrow k$ and $j\longleftrightarrow l$ as well all
we have used the symmetry properties of $\boldsymbol{g}_{\boldsymbol{i}}$.
\begin{table}
\begin{tabular}{|c|c|c|c|}
\hline 
Term & Swapping indices & sign $\boldsymbol{g}_{\boldsymbol{i}}$ & Final term\tabularnewline
\hline 
$\boldsymbol{g}_{\boldsymbol{i}}x_{ik}x_{\beta j}\delta_{l\alpha}$ & $k\longleftrightarrow j$ $\left(k\longleftrightarrow l\right)$ & $-$ $\left(-\right)$ & $\boldsymbol{g}_{\boldsymbol{i}}x_{ij}x_{\beta l}\delta_{k\alpha}$\tabularnewline
\hline 
$\boldsymbol{g}_{\boldsymbol{i}}\delta_{\beta l}x_{ik}x_{j\alpha}$ & $k\longleftrightarrow j$  & $-$  & $-\boldsymbol{g}_{\boldsymbol{i}}x_{ij}x_{k\alpha}\delta_{\beta l}$\tabularnewline
\hline 
$-2\boldsymbol{g}_{\boldsymbol{i}}x_{il}x_{\beta j}\delta_{k\alpha}$ & $l\longleftrightarrow j$  & $-$  & $2\boldsymbol{g}_{\boldsymbol{i}}x_{ij}x_{\beta l}\delta_{k\alpha}$\tabularnewline
\hline 
$2\boldsymbol{g}_{\boldsymbol{i}}x_{kj}x_{i\alpha}\delta_{\beta l}$ & $k\longleftrightarrow i$  & $-$  & $-2\boldsymbol{g}_{\boldsymbol{i}}x_{ij}x_{k\alpha}\delta_{\beta l}$\tabularnewline
\hline 
$-\boldsymbol{g}_{\boldsymbol{i}}x_{lj}x_{i\alpha}\delta_{k\beta}$ & $l\longleftrightarrow i$ $\left(k\longleftrightarrow l\right)$ & $-$ $\left(-\right)$ & $-\boldsymbol{g}_{\boldsymbol{i}}x_{ij}x_{k\alpha}\delta_{l\beta}$\tabularnewline
\hline 
$\boldsymbol{g}_{\boldsymbol{i}}x_{i\beta}x_{lj}\delta_{k\alpha}$ & $l\longleftrightarrow i$  & $-$  & $-\boldsymbol{g}_{\boldsymbol{i}}x_{l\beta}x_{ij}\delta_{k\alpha}$\tabularnewline
\hline 
\end{tabular}

\caption{Transformation table that indicates the corresponding exchange of
indices, performing in each term of Eq. (\ref{eq:DerDxmunuNS}). The
brackets $\left(\ldots\right)$ indicate a second swapping that is
performed for that particular term. \label{tab:TransfTable}}
\end{table}
The second order derivative of the diffusion term is calculated
using the following expression given in Eq. (\ref{eq:DermunuDifTerm}):
\begin{eqnarray}
\partial_{\bm{\mu}}D^{\bm{\alpha}\bm{\mu}} & = & -16\left(3-2M\right)\boldsymbol{g}_{\boldsymbol{i}}x_{kl}\Im X_{ij}^{\alpha\beta}.
\end{eqnarray}
Using the product rule of the derivative we prove $\frac{\partial\Im X_{ij}^{\alpha\beta}}{\partial x_{\alpha\beta}}=0$.
Implementation of this gives,
\begin{eqnarray}
\partial_{\bm{\alpha}}\left(x_{kl}\Im X_{ij}^{\alpha\beta}\right) & = & \left(\delta_{k\alpha}\delta_{l\beta}-\delta_{k\beta}\delta_{l\alpha}\right)\Im X_{ij}^{\alpha\beta}\nonumber \\
 & = & \left(\Im X_{ij}^{kl}-\Im X_{ij}^{lk}\right)\nonumber \\
 & = & 0.\label{eq:DerxklIXijab}
\end{eqnarray}
This last result follows since we are considering that $i\neq j\neq k\neq l$
, therefore:
\begin{eqnarray}
\Im X_{ij}^{kl} & = & -\left(x_{ik}\delta_{lj}-\delta_{ik}x_{lj}\right)=0.
\end{eqnarray}
All other permutations of indices $i,$ $j$, $k$ and $l$ for $X_{ij}^{kl}$,
gives the same result. Therefore we have proved that:
\begin{eqnarray}
\partial_{\bm{\bm{\alpha}}}\partial_{\bm{\bm{\mu}}}D^{\bm{\alpha}\bm{\mu}} & = & -\frac{32}{\hbar}\left(3-2M\right)\frac{\partial}{\partial x_{\alpha\beta}}x_{kl}\Im X_{ij}^{\alpha\beta}\nonumber \\
 & = & 0.\label{eq:SODerDifT}
\end{eqnarray}

\subsection{Derivatives of the drift term}

In order to calculate the first order derivative of the drift term
$A^{\bm{\alpha}}\equiv A^{\left(\alpha\beta\right)},$ we consider
the expression given in Eq. (\ref{eq:DriftTerm}) as well as Eq. (\ref{eq:driftfirst}).
So
\begin{equation}
\partial_{\bm{\alpha}}A^{\bm{\alpha}}=\partial_{\bm{\alpha}}\bar{A}^{\bm{\alpha}}+\partial_{\bm{\bm{\alpha}}}\partial_{\bm{\bm{\mu}}}D^{\bm{\alpha}\bm{\mu}}.
\end{equation}

As the second order derivative of the diffusion term is zero from
Eq. (\ref{eq:DerxklIXijab}), the second term of the above expression
is zero. Utilizing the result Eq. (\ref{eq:Der}), one can obtain
that:
\begin{align}
\partial_{\bm{\alpha}}\bar{A}^{\bm{\alpha}} & =4\partial_{\bm{\alpha}}\left[\Im X_{ij}^{\bm{\alpha}}\left(3\boldsymbol{g}_{\boldsymbol{i}}x_{kl}^{+}-t_{ij}\right)\right]\nonumber \\
 & =12\partial_{\bm{\alpha}}\left(\Im X_{ij}^{\bm{\alpha}}\boldsymbol{g}_{\boldsymbol{i}}x_{kl}^{+}\right)-4\partial_{\bm{\alpha}}\left(\Im X_{ij}^{\bm{\alpha}}t_{ij}\right).
\end{align}

From Eq. (\ref{eq:DerxklIXijab}) and Eq. (\ref{eq:Der}) one know
that $\partial_{\bm{\alpha}}\left(x_{kl}^{+}\Im X_{ij}^{\bm{\alpha}}\right)=0$
and $\partial_{\bm{\alpha}}\Im X_{ij}^{\bm{\alpha}}=0,$ so one can
obtain that the first order derivative of the drift term is zero.
\begin{eqnarray}
\frac{\partial A^{\left(\alpha\beta\right)}}{\partial x_{\alpha\beta}} & = & 0.
\end{eqnarray}

This gives the result, in a more compact form, that $\partial_{\bm{\alpha}}A^{\bm{\alpha}}=0.$

\end{document}